\documentclass[aps,prd,superscriptaddress,notitlepage,showpacs,showkeys,10pt]{revtex4-1}

\usepackage{graphicx}
\usepackage{amsmath}
\usepackage{amssymb}
\usepackage{color}
\usepackage{bm}

\newcommand{\sign}[1]{\mbox{sgn}\left({#1}\right)}

\definecolor{purple}{rgb}{0.8,0,0.6}
\definecolor{darkgreen}{rgb}{0.00,0.6,0.00}

\begin{document}
\title{Second-order chiral kinetic theory: Chiral magnetic and pseudomagnetic waves}
\date{\today}

\author{E.~V.~Gorbar}
\affiliation{Department of Physics, Taras Shevchenko National Kiev University, Kiev, 03680, Ukraine}
\affiliation{Bogolyubov Institute for Theoretical Physics, Kiev, 03680, Ukraine}

\author{V.~A.~Miransky}
\affiliation{Department of Applied Mathematics, Western University, London, Ontario, Canada N6A 5B7}
\affiliation{Department of Physics and Astronomy, Western University, London, Ontario, Canada N6A 5B7}

\author{I. A.~Shovkovy}
\affiliation{College of Integrative Sciences and Arts, Arizona State University, Mesa, Arizona 85212, USA}
\affiliation{Department of Physics, Arizona State University, Tempe, Arizona 85287, USA}

\author{P.~O.~Sukhachov}
\affiliation{Department of Applied Mathematics, Western University, London, Ontario N6A 5B7, Canada}

\begin{abstract}
The consistent chiral kinetic theory accurate to the second order in electromagnetic and pseudoelectromagnetic
fields is derived for a relativistic matter with two Weyl fermions. By making use of such a framework,
the properties of longitudinal collective excitations, which include both chiral magnetic and chiral pseudomagnetic
waves, are studied. It is shown that the proper treatment of dynamical electromagnetism transforms these gapless
waves into chiral (pseudo)magnetic plasmons, whose Langmuir (plasma) gap receives corrections quadratic in both
magnetic and pseudomagnetic fields.
\end{abstract}



\maketitle

\section{Introduction}
\label{sec:introduction}

A kinetic theory is a common framework for studying a wide range of physical properties in condensed
matter materials, nuclear physics, and cosmology \cite{Landau:t10,Krall}. A relativistic version of such a
theory is invaluable in studies of numerous relativistic forms of matter, e.g., realized
in the primordial plasma of the early universe \cite{Kronberg,Durrer}, relativistic heavy-ion collisions
\cite{Kharzeev:2008-Nucl,Kharzeev:2016}, compact stars \cite{Kouveliotou:1999}, as well as the Dirac and Weyl materials.
Following the first theoretical predictions in Refs.~\cite{Wang,Weng,Weng:2014}, the realization of the three-dimensional
(3D) Dirac semimetal phase in A$_3$Bi ($\mathrm{A=Na,K,Rb}$), Cd$_3$As$_2$, and ZrTe$_5$ was confirmed experimentally \cite{Borisenko,Neupane,Liu,Xiong:2015,Li-Wang:2015,Li-He:2015,Li:2014bha,Zheng:2016}. While the Weyl semimetal phase was first predicted theoretically to be realized in pyrochlore
iridates \cite{Savrasov}, it was discovered later in such compounds as $\mathrm{TaAs}$, $\mathrm{TaP}$,
$\mathrm{NbAs}$, $\mathrm{NbP}$, $\mathrm{Mo_xW_{1-x}Te}$, and $\mathrm{YbMnBi_2}$ \cite{Weng-Fang:2015,Qian,Huang:2015eia,Bian,Huang:2015Nature,Zhang:2016,Cava,Belopolski}. The
above-mentioned forms of matter are often made of (approximately) massless fermionic particles that carry a well-defined chirality.
Therefore, the chirality could be treated as a natural additional degree of freedom in a relativistic plasma.
It should be noted, however, that the conservation of the chiral charge is violated in the presence of external
electromagnetic fields. This is a consequence of the celebrated chiral (triangle) anomaly \cite{ABJ}. Nevertheless, this anomaly can be incorporated exactly in the chiral kinetic theory \cite{Son:2012wh,Stephanov:2012ki}.
The only limitation of the current formulation of the corresponding theory is that it is valid only to the linear order in
electromagnetic fields.

It important to note that the chiral anomaly affects magnetotransport in Weyl and Dirac semimetals in a nontrivial way.
Indeed, as was first shown by Nielsen and Ninomiya \cite{Nielsen}, the longitudinal (with respect to the
direction of an external magnetic field) magnetoresistivity in Weyl semimetals decreases with the growth of the magnetic
field. Physically this phenomenon, which is usually called negative magnetoresistivity, relies on the presence of the lowest Landau level (LLL) states around each Weyl node \cite{Nielsen,Son}. Then, since the LLL density of
states grows linearly with a magnetic field, the conductivity increases too. The phenomenon of negative magnetoresistivity is extensively studied both theoretically \cite{Nielsen,Son,Aji:2012,Kim:2013dia,Gorbar:2013dha,Burkov:2015} and experimentally \cite{Li:2014bha,Xiong:2015,Li-He:2015,Li-Wang:2015,Huang:2015eia,Zhang:2016}.

The presence of a chiral chemical potential $\mu_5$, which quantifies a chiral asymmetry in a relativistic matter,
significantly enriches the properties and types of collective excitations. The investigation of the corresponding
features began in Refs.~\cite{Kharzeev:2010gd,Akamatsu:2013,Stephanov:2015}. The authors of Ref.~\cite{Kharzeev:2010gd}
showed that the triangle anomaly implies the existence of a novel type of collective excitations which stems from the coupling between the
density waves of the electric and chiral charges and is known as the chiral magnetic wave (CMW).
In this connection, let us also
mention that, as was shown in Ref.~\cite{Akamatsu:2013}, a nonzero chiral chemical potential lifts the degeneracy
of the transverse plasma modes, but leaves the longitudinal mode intact. Such a conclusion agrees with a refined
analysis in the consistent chiral kinetic theory \cite{Gorbar:2016ygi,Gorbar:2016sey}.

Dirac and Weyl materials provide additional opportunities to explore the role of chirality. For example, they allow
for a simple realization of axial electric (or pseudoelectric) $\mathbf{E}_5$ and/or
axial magnetic (or pseudomagnetic) $\mathbf{B}_5$ fields. These pseudoelectromagnetic fields
act on fermions as ordinary electromagnetic fields, but their sign depends on the fermion chirality. In Weyl
and Dirac materials, one can induce a background pseudomagnetic field $\mathbf{B}_5$ by various kinds of static
strains \cite{Zubkov:2015,Cortijo:2016yph,Pikulin:2016,Cortijo:2016,Grushin-Vishwanath:2016,Liu-Pikulin:2016}.
Note that, unlike an ordinary magnetic field $\mathbf{B}$, a pseudomagnetic one does not break the time-reversal symmetry.
In essence, this is due to the fact that Weyl nodes in condensed matter materials always come in pairs of opposite
chirality \cite{Nielsen:1980rz}. A pseudoelectric $\mathbf{E}_5$ can be induced by time-dependent strains \cite{Pikulin:2016}.

Recently, we showed \cite{Gorbar:2016ygi,Gorbar:2016sey} that the plasmons in a relativistic matter in constant
magnetic and pseudomagnetic fields are, in fact, {\em chiral} (pseudo)magnetic plasmons. Their chiral nature
is manifested by oscillations of a chiral charge density, which are absent for ordinary electromagnetic plasmons.
It is also worth noting that, in the presence of magnetic fields, the quasiparticle plasma in uncompensated metals
supports a special type of collective excitations known as helicons. These transverse low-energy gapless excitations
propagate along the background magnetic field. According to the studies in Ref.~\cite{Pellegrino}, the dispersion
relation of helicons encodes information on the chiral shift parameter $\mathbf{b}$, which defines the
momentum space separation of Weyl nodes. Further, by using the formalism
of the consistent chiral kinetic theory, it was shown in Ref.~\cite{Gorbar:2016vvg} that pseudomagnetic fields allow
for a new type of helicons, which we called pseudomagnetic helicons.

Although the chiral kinetic theory is quite successful in the description of various processes in relativistic plasma,
many physical phenomena require its formulation accurate to the second order in electromagnetic fields. Recently,
using the wave-packet semiclassical approach \cite{Sundaram:1999zz} (for a review, see Ref.~\cite{Xiao:2009rm}),
all the necessary ingredients for
such a formulation were provided in Refs.~\cite{Gao:2014,Gao:2015} for condensed matter systems with a
general band structure. In the present paper, we derive the explicit expressions for the consistent chiral kinetic
theory valid to the second order in electromagnetic, as well as pseudoelectromagnetic fields for a simple realization of relativistic matter in a Weyl
material with a single pair of Weyl nodes.

The effects of dynamical electromagnetism for the chiral magnetic and chiral pseudomagnetic waves were
studied in Ref.~\cite{Gorbar:2016sey} in the consistent chiral kinetic theory valid to the linear order
in electromagnetic and pseudoelectromagnetic fields. It was found that such excitations are chiral
(pseudo)magnetic plasmons with the field-independent Langmuir (plasma) gap. While the background magnetic field
does not affect at all the dispersion relation in the linear order, the background
pseudomagnetic field contributes only to the term linear in momentum. However, according to the
general arguments in Ref.~\cite{Kharzeev:2010gd}, the dispersion relation of the CMW with the effects of dynamical
electromagnetism included should have the form $\omega^2=\Omega^2_e+(v^2_{\chi}+c^2_{\rm eff})\mathbf{k}^2$,
where $v_{\chi}\propto eB$. Since $v^2_{\chi}$ is proportional to the square of the magnetic field,
this result cannot be reliably reproduced in the conventional first-order chiral kinetic theory \cite{Son:2012wh,Stephanov:2012ki}, unless the theory
is generalized to the second order in electromagnetic fields. In essence, this is one of the main motivations
for this paper.

The paper is organized as follows. The consistent chiral kinetic theory for a Weyl material with two
Weyl fermions valid to the second order in electromagnetic and pseudoelectromagnetic fields is
formulated in Sec.~\ref{sec:second-CKT-Berry}. The longitudinal collective excitations propagating
along background magnetic and pseudomagnetic fields are considered in Sec.~\ref{sec:second-collective}.
The summary of the main results is given in Sec.~\ref{sec:Summary-Discussions}. Some useful technical
results and formulas are presented in Appendixes \ref{sec:App-Corrections} and \ref{sec:App-Polarization}.

\section{Second-order chiral kinetic theory}
\label{sec:second-CKT-Berry}

Let us start by discussing the form of the second order in electromagnetic and pseudoelectromagnetic field corrections to the quasiparticle energy and
velocity, as well as the leading-order corrections to the Berry curvature. (Indeed, the first-order corrections to the Berry curvature are sufficient because in the chiral kinetic theory it always couples to electromagnetic fields \cite{Gao:2014,Gao:2015}.)
Our starting point is the
Hamiltonian for a single Weyl fermion
\begin{equation}
H=\lambda v_F (\mathbf{p}\cdot\bm{\sigma}),
\label{second-def-H}
\end{equation}
where $\lambda=\pm$ is chirality, $v_F$ is the Fermi velocity, $\mathbf{p}$ is a momentum, and $\bm{\sigma}$ are the Pauli matrices. In the absence of external electromagnetic fields,
the quasiparticle energy is $\epsilon^{(0)}_{\eta} =\eta  v_F p$, where $\eta$ labels the quasiparticles in the upper
($\eta=+1$) or lower ($\eta=-1$) band. (Note that in relativistic language, the quasiparticles in the upper/lower band correspond
to the particles/antiparticles.)
In the absence of background fields, the Berry curvature in the reciprocal (momentum) space is given by \cite{Berry:1984}
\begin{equation}
\mathbf{\Omega}_{\lambda}^{(0)}= \lambda \eta  \hbar \frac{\mathbf{p}}{2p^3}.
\label{def-Omega-0}
\end{equation}
In this study we will assume that the Weyl fermions are in the effective electric and magnetic
fields,
\begin{eqnarray}
\mathbf{E}_{\lambda}=\mathbf{E}+\lambda\mathbf{E}_{5}, \qquad \mathbf{B}_{\lambda}=\mathbf{B}+\lambda\mathbf{B}_{5},
\label{second-CKT-fields}
\end{eqnarray}
where $\mathbf{E}$ and $\mathbf{B}$ are the usual electric and magnetic fields, while $\mathbf{E}_{5}$ and
$\mathbf{B}_{5}$ are pseudoelectromagnetic fields induced by strains. Henceforth, we will
assume that there are no dynamical strains in the sample and, therefore, the pseudoelectric field vanishes,
$\mathbf{E}_5=\mathbf{0}$. A pseudomagnetic field can
be generated, e.g., either by applying a static torsion \cite{Pikulin:2016} or by bending \cite{Liu-Pikulin:2016} the
sample. The estimated values of the field could be somewhere in range from $B_5\approx0.3$ to
$15~\mbox{T}$.

In the presence of background fields (\ref{second-CKT-fields}), the quasiparticle dispersion relation should
receive a correction due to the interaction of the quasiparticle's magnetic moment with the effective magnetic field. At linear
order, the corresponding contribution to the energy is proportional to the scalar product of the field and the
Berry curvature. To the second order in electromagnetic fields \cite{Gao:2014,Gao:2015}, there are additional
corrections to the quasiparticle energy, as well as to the Berry curvature in Eq.~(\ref{def-Omega-0}).
The explicit expressions for the corrections to the Berry curvature, the quasiparticle dispersion
relation, and the quasiparticle velocity are derived in Appendix~\ref{sec:App-Corrections}. In particular,
the final results for the Berry curvature and the quasiparticle energy read as
\begin{eqnarray}
\mathbf{\Omega}_{\lambda} &\equiv&  \mathbf{\Omega}_{\lambda}^{(0)} +\mathbf{\Omega}_{\lambda}^{(1)} =
\lambda \eta  \hbar \frac{\hat{\mathbf{p}}}{2p^2}
+\frac{e \hbar^2}{4p^4} \left\{\frac{2}{c}\hat{\mathbf{p}}(\hat{\mathbf{p}}\cdot\mathbf{B}_{\lambda})
-\frac{1}{c}\mathbf{B}_{\lambda} +\frac{2\eta }{v_F}[\mathbf{E}_{\lambda}\times\hat{\mathbf{p}}]\right\} ,
\label{def-Omega-1} \\
\epsilon_{\mathbf{p}} &=& \eta  v_Fp
-\lambda\frac{e\hbar v_F }{2cp} (\mathbf{B}_{\lambda}\cdot\hat{\mathbf{p}}) + \frac{e^2 \hbar^2}{4cp^3} \left\{\frac{\eta  v_F}{4c}
\left[2 B_{\lambda}^2-(\mathbf{B}_{\lambda}\cdot\hat{\mathbf{p}})^2\right] - (\mathbf{B}_{\lambda}\cdot[\mathbf{E}_{\lambda}\times
\hat{\mathbf{p}}])\right\},
\label{second-energy-2}
\end{eqnarray}
where $\hat{\mathbf{p}}=\mathbf{p}/|\mathbf{p}|$. It is also straightforward to derive the explicit expression for the
quasiparticle velocity, $\mathbf{v}=\partial \epsilon_{\mathbf{p}}/{\partial \mathbf{p}}$ [see Eq.~(\ref{second-energy-v})
in Appendix~\ref{sec:App-Corrections}]. Having determined the second-order
corrections to the quasiparticle energy and the leading-order corrections to the Berry curvature, we can now
formulate the consistent chiral kinetic theory valid to the second order in (pseudo)electromagnetic fields.
Before proceeding to the kinetic equation, however, it may be instructive to point out that the corrected
expression (\ref{def-Omega-1}) for the Berry curvature $\mathbf{\Omega}_{\lambda}$ corresponds to the same unit topological
charge as the original configuration in Eq.~(\ref{def-Omega-0}). This follows from the fact that
\begin{equation}
\int\frac{d^3 p}{(2\pi)^3} \left(\frac{\partial }{\partial \mathbf{p}} \cdot \mathbf{\Omega}_{\lambda}^{(1)} \right)=0.
\label{second-def-Berry-curv-div-ee}
\end{equation}
Thus, although the Berry curvature and equations of motion are corrected, the chiral anomaly relation
retains its canonical form in the second-order formulation of the chiral kinetic theory.

In the phase space, the one-particle distribution functions $f_{\lambda}(\mathbf{p},\mathbf{r})$ for the
right- ($\lambda=+$) and left-handed ($\lambda=-$) fermions satisfy the following kinetic equation:
\begin{equation}
\frac{\partial f_{\lambda}}{\partial t}+\dot{\mathbf{p}}\cdot\frac{\partial f_{\lambda}}{\partial
\mathbf{p}}+\dot{\mathbf{r}}\cdot\frac{\partial f_{\lambda}}{\partial \mathbf{r}}=I_{\rm coll}(f_{\lambda}),
\label{second-CKT-kinetic-equations-general}
\end{equation}
where the term on the right-hand side is the collision integral and, for the sake of brevity, we dropped the arguments in $f_{\lambda}$.
In what follows, we will consider the collisionless limit. Therefore, $I_{\rm coll}(f_{\lambda}) \equiv 0$.

The equations of motion for quasiparticles to the second order in electromagnetic
fields were derived in Ref.~\cite{Gao:2014}. Surprisingly, their general form is the same as in the first-order theory, i.e.,
\begin{eqnarray}
\dot{\mathbf{r}} &=& \mathbf{v}+(\dot{\mathbf{p}}\times\mathbf{\Omega}_{\lambda}),
\label{second-CKT-equations-motion-x}\\
\dot{\mathbf{p}} &=& e\tilde{\mathbf{E}}_{\lambda} +\frac{e}{c}(\dot{\mathbf{r}}\times\mathbf{B}_{\lambda}),
\label{second-CKT-equations-motion-p}
\end{eqnarray}
where $\tilde{\mathbf{E}}_{\lambda} = \mathbf{E}_{\lambda}-(1/e)\partial \epsilon_{\mathbf{p}}/\partial \mathbf{r}$.
However, it should be emphasized that the expressions for the Berry curvature $\mathbf{\Omega}_{\lambda}$ and the
quasiparticle energy $\epsilon_{\mathbf{p}}$ are more complicated and are given by Eqs.~(\ref{def-Omega-1}) and (\ref{second-energy-2}),
respectively.

After making use of Eq.~(\ref{second-CKT-equations-motion-x}) and (\ref{second-CKT-equations-motion-p}), the
chiral kinetic equation (\ref{second-CKT-kinetic-equations-general}) can be rewritten in the usual form,
\begin{equation}
\frac{\partial f_{\lambda}}{\partial t}+\frac{1}{1+\frac{e}{c}(\mathbf{B}_{\lambda}\cdot\mathbf{\Omega}_{\lambda})}
\left[\Big(e\tilde{\mathbf{E}}_{\lambda}
+\frac{e}{c}(\mathbf{v}\times \mathbf{B}_{\lambda})
+\frac{e^2}{c}(\tilde{\mathbf{E}}_{\lambda}\cdot\mathbf{B}_{\lambda})\mathbf{\Omega}_{\lambda}\Big)\cdot\frac{\partial
f_{\lambda}}{\partial \mathbf{p}}
+\Big(\mathbf{v}+e(\tilde{\mathbf{E}}_{\lambda}\times\mathbf{\Omega}_{\lambda})
+\frac{e}{c}(\mathbf{v}\cdot\mathbf{\Omega}_{\lambda})\mathbf{B}_{\lambda}\Big)\cdot\frac{\partial f_{\lambda}}{\partial \mathbf{r}}\right]=0,
\label{second-CKT-kinetic-equation}
\end{equation}
where the factor $1/[1+e(\mathbf{B}_{\lambda}\cdot\mathbf{\Omega}_{\lambda})/c]$ accounts for the correct definition
of the phase-space volume that satisfies the Liouville's theorem \cite{Xiao,Duval}. This form of the kinetic equation is
identical to that in the first-order chiral kinetic theory. One should keep in mind, however, that the expressions for
the Berry curvature, the quasiparticle energy, and the quasiparticle velocity include the additional corrections discussed
above.

The formal definitions of the fermion charge and current densities are given by the same expressions as in the
first-order chiral kinetic theory, i.e.,
\begin{equation}
\rho_{\lambda} =\sum_{\eta=\pm} \eta\,e\int\frac{d^3p}{(2\pi \hbar)^3}
\left[1+\frac{e}{c}\left(\mathbf{B}_{\lambda}\cdot\mathbf{\Omega}_{\lambda}\right)\right]f_{\lambda}
\label{second-CKT-charge-density}
\end{equation}
and
\begin{equation}
\mathbf{j}_{\lambda} =\sum_{\eta=\pm} \eta\, e\int\frac{d^3p}{(2\pi \hbar)^3}\left[\mathbf{v}
+e(\tilde{\mathbf{E}}_{\lambda}\times\mathbf{\Omega}_{\lambda})
+\frac{e}{c}(\mathbf{v} \cdot\mathbf{\Omega}_{\lambda})\mathbf{B}_{\lambda}
\right]\,f_{\lambda}
+\sum_{\eta=\pm} \eta \,e \int\frac{d^3p}{(2\pi \hbar)^3}\left(\partial_{\mathbf{r}}\times f_{\lambda}\epsilon_{\mathbf{p}}\mathbf{\Omega}_{\lambda} \right)
+O(B_{\lambda}^3).
\label{second-CKT-electric-current-b}
\end{equation}
Note that the last term in Eq.~(\ref{second-CKT-electric-current-b}) is the magnetization current.

In the consistent chiral kinetic theory \cite{Gorbar:2016ygi,Gorbar:2016sey}, an additional topological contribution
\cite{Bardeen,Landsteiner:2013sja,Landsteiner:2016} to the four-current density, $\delta j^{\mu} \equiv (c\delta \rho,
\delta \mathbf{j})$, is required,
\begin{equation}
\delta j^{\mu} =  \frac{e^3}{4\pi^2 \hbar^2 c} \epsilon^{\mu \nu \rho \lambda} A_{\nu}^5 F_{\rho \lambda},
\label{second-consistent-def-0}
\end{equation}
where $A_{\nu}^5=b_{\nu}+\tilde{A}^5_{\nu}$ is the axial potential. Unlike the electromagnetic potential
$A_{\nu}$, the axial potential is an observable quantity. Indeed, in Weyl materials, $b_0$ and $\mathbf{b}$
describe the separations between the Weyl nodes in energy and momentum, respectively. Strain-induced
axial (or pseudoelectromagnetic) fields are described by $\tilde{A}_{\nu}^5$, which is directly related to the
deformation tensor \cite{Zubkov:2015,Cortijo:2016yph,Cortijo:2016,Grushin-Vishwanath:2016,Pikulin:2016,Liu-Pikulin:2016}.
It is easy to check that contrary to the covariant electric current $j^{\mu} \equiv (c \rho, \mathbf{j})$,
the consistent one,
\begin{equation}
J^{\nu} \equiv (c\rho+c\delta \rho, \mathbf{j}+\delta \mathbf{j}),
\label{second-consistent-4-current}
\end{equation}
is nonanomalous, $\partial_{\nu} J^{\nu}=0$, or, in other words, the electric charge is locally conserved in the
presence of pseudoelectromagnetic fields (for a detailed discussion, see Refs.~\cite{Gorbar:2016ygi,Gorbar:2016sey}).
Note that we introduced the following short-hand notations in Eq.~(\ref{second-consistent-4-current}):
\begin{equation}
\rho=\sum_{\lambda=\pm}\,\rho_{\lambda}, \quad\quad
\mathbf{j}=\sum_{\lambda=\pm}\,\mathbf{j}_{\lambda},
\label{second-consistent-charge-current}
\end{equation}
and used the component form of the topological contribution in Eq.~(\ref{second-consistent-def-0}), i.e.,
\begin{eqnarray}
\delta \rho &=&\frac{e^3}{2\pi^2 \hbar^2c^2}\,(\mathbf{b}\cdot\mathbf{B}),
\label{second-consistent-charge-density}
\\
\delta \mathbf{j} &=&\frac{e^3}{2\pi^2 \hbar^2 c}\,b_0 \,\mathbf{B} - \frac{e^3}{2\pi^2 \hbar^2 c}\,(\mathbf{b}\times\mathbf{E}).
\label{second-consistent-current-density}
\end{eqnarray}
Here, we assumed that $\tilde{\mathbf{A}}^5$ is negligible compared to the chiral shift $\mathbf{b}$ and set
$\tilde{A}^5_0=0$ in accordance with our assumption that a pseudoelectric field is absent.

\section{Electromagnetic collective modes}
\label{sec:second-collective}

\subsection{General consideration}
\label{sec:second-collective-general-eq}

In this section, using of the formalism of the second-order consistent chiral kinetic theory, we determine the dispersion relations of the collective excitations in strained Weyl materials in the
presence of a constant background field $\mathbf{B}_{0,\lambda}\equiv \mathbf{B}_0+\lambda \mathbf{B}_{0,5}$. For simplicity, we
assume that the ordinary magnetic field $\mathbf{B}_0$ and the strain-induced pseudomagnetic field $\mathbf{B}_{0,5}$
are parallel to each other. In addition to the background fields $\mathbf{B}_0$ and $\mathbf{B}_{0,5}$,
oscillating electromagnetic fields $\mathbf{E}^{\prime}$ and $\mathbf{B}^{\prime}$ will be induced by collective
modes. In principle, one might speculate that $\mathbf{E}^{\prime}$ and $\mathbf{B}^{\prime}$ could in turn drive dynamical
deformations of the Weyl material and, thus, generate oscillating pseudomagnetic fields $\mathbf{E}_5^{\prime}$
and $\mathbf{B}_5^{\prime}$. The latter, however, are extremely weak \cite{Gorbar:2016ygi,Gorbar:2016sey} and
will be neglected in our analysis.

Our consideration of electromagnetic collective modes uses the standard approach of physical kinetics
\cite{Krall,Landau:t10}, but generalized to account for the Berry curvature, the pseudomagnetic field, and
the topological current correction. As usual, we seek the solutions in the form of plain waves
\begin{equation}
\mathbf{E}^{\prime} = \mathbf{E} e^{-i\omega t+i\mathbf{k}\cdot\mathbf{r}} ,
\qquad
\mathbf{B}^{\prime} = \mathbf{B} e^{-i\omega t+i\mathbf{k}\cdot\mathbf{r}} ,
\label{oscillating-fields}
\end{equation}
with frequency $\omega$ and wave vector $\mathbf{k}$. The Maxwell's equations imply
that $\mathbf{B}^{\prime} = c(\mathbf{k}\times \mathbf{E}^{\prime})/\omega$ and
\begin{eqnarray}
\mathbf{k}\left(\mathbf{k}\cdot \mathbf{E}^{\prime} \right) - k^2 \mathbf{E}^{\prime}
=-\frac{\omega^2}{c^2}\left(n_0^2\mathbf{E}^{\prime}+4\pi \mathbf{P}^{\prime}\right).
\label{second-collective-B-tensor-spectrum-EM-0}
\end{eqnarray}
Here, $\mathbf{P}^{\prime}$ denotes the polarization vector and $n_0$ is the background refractive index
of the material. For the Dirac semimetal Cd$_3$As$_2$, e.g., the latter is $n_0\approx6$ \cite{Freyland}.
In order to simplify the analysis, we will neglect the dependence of the refractive index on the frequency.

By introducing the electric
susceptibility tensor $\chi^{mn}$ (where $m,n=1,2,3$ denote spatial components), the polarization
vector takes the form
\begin{equation}
P^{\prime m}= i\frac{J^{\prime m}}{\omega}=\chi^{mn}E^{\prime n},
\label{second-polarization-tensor}
\end{equation}
where $J^{\prime m}$ is defined by Eq.~(\ref{second-consistent-4-current}). Then,
Eq.~(\ref{second-collective-B-tensor-spectrum-EM-0}) implies
\begin{eqnarray}
\left(n_0^2\omega^2- c^2k^2 \right)\delta^{mn}E^n= -c^2 k^m k^nE^n -4\pi\omega^2\chi^{mn}E^n.
\label{second-collective-B-tensor-spectrum-EM-1}
\end{eqnarray}
The above equation admits nontrivial solutions only if the corresponding determinant vanishes, i.e.,
\begin{equation}
\mbox{det}\left[\left(n_0^2\omega^2- c^2k^2 \right)\delta^{mn} + c^2 k^m k^n + 4\pi\omega^2\chi^{mn}\right]=0.
\label{second-collective-B-tensor-dispersion-relation-general}
\end{equation}
This characteristic equation defines the dispersion relation of electromagnetic collective modes.

In order to determine the susceptibility tensor $\chi^{mn}$ in the consistent chiral kinetic theory,
we choose the usual ansatz for the distribution function in the form $f_{\lambda}=f_{\lambda}^{\rm (eq)}+\delta f_{\lambda}$,
where $f_{\lambda}^{\rm (eq)}$ is the equilibrium distribution function for the electrons (holes) in the upper (lower) band given by
\begin{equation}
f_{\lambda}^{\rm (eq)}=\frac{1}{e^{\eta(\epsilon_{\mathbf{p}}-\mu_{\lambda})/T}+1}.
\label{second-CKT-equilibrium-function}
\end{equation}
Here, $T$ is the temperature and $\mu_{\lambda}=\mu+\lambda\mu_5$ is the chemical potential for the
fermions of chirality $\lambda$. The latter is conveniently expressed in terms of the fermion-number
chemical potential $\mu$, as well as chiral-charge chemical potential $\mu_5$. Note that we set the
Boltzmann constant to unity $k_B=1$. It should be emphasized that the form of the equilibrium
distribution function (\ref{second-CKT-equilibrium-function}) is valid for quasiparticles in both the lower
and upper bands.

Due to the oscillating $\mathbf{E}^{\prime}$ and $\mathbf{B}^{\prime}$ fields defined in Eq.~(\ref{oscillating-fields}), the
corresponding perturbation
to the equilibrium distribution function is also of the plane wave form, i.e.,
\begin{equation}
\delta f_{\lambda} = f_{\lambda}^{(1)} e^{-i\omega t+i\mathbf{k}\cdot\mathbf{r}} .
\end{equation}
In the first order in oscillating electromagnetic fields, the chiral kinetic equation
(\ref{second-CKT-kinetic-equation}) gives
\begin{eqnarray}
&&-i\omega \delta f_{\lambda} +\frac{1}{1+\kappa_{\lambda}}
\left[e\tilde{\mathbf{E}}^{\prime}_{\lambda}
+\frac{e}{c}(\mathbf{v}_0\times \mathbf{B}^{\prime}) +\frac{e}{c}(\mathbf{v}^{\prime}\times \mathbf{B}_{0,\lambda})
+\frac{e^2}{c}(\tilde{\mathbf{E}}_{\lambda}^{\prime}\cdot\mathbf{B}_{0,\lambda})\mathbf{\Omega}_{0,\lambda} \right]\cdot\mathbf{v}_{0}
\frac{\partial f_{\lambda}^{\rm (eq)}}{\partial \epsilon_\mathbf{p}} \nonumber\\
&&+\frac{1}{1+\kappa_{\lambda}} \frac{e}{c}(\mathbf{v}_0\times \mathbf{B}_{0,\lambda})\cdot
\frac{\partial \delta f_{\lambda}}{\partial \mathbf{p}}
+\frac{i}{1+\kappa_{\lambda}}\left[(\mathbf{v}_0\cdot\mathbf{k})
+\frac{e}{c}(\mathbf{v}_0\cdot\mathbf{\Omega}_{0,\lambda})(\mathbf{B}_{0,\lambda}\cdot\mathbf{k})\right]\delta f_{\lambda}=0,
\label{second-collective-kinetic-equation}
\end{eqnarray}
where
\begin{eqnarray}
\label{second-collective-formulas-be}
&&\mathbf{v}_0\equiv\mathbf{v}\Big|_{\mathbf{B}_{\lambda}=\mathbf{B}_{0,\lambda}, \mathbf{E}_{\lambda}=0},\quad\quad
\mathbf{\Omega}_{0,\lambda}\equiv\mathbf{\Omega}_{\lambda}\Big|_{\mathbf{B}_{\lambda}=\mathbf{B}_{0,\lambda}, \mathbf{E}_{\lambda}=0},\\
&&\tilde{\mathbf{E}}_{\lambda}^{\prime} = \mathbf{E}^{\prime} +i\frac{\lambda \hbar v_F}{2cp}
\mathbf{k}(\hat{\mathbf{p}}\cdot\mathbf{B}^{\prime}) +i\frac{\eta  e v_F \hbar^2}{8c^2p^3} \mathbf{k}(\mathbf{B}_{0,\lambda}\cdot
\hat{\mathbf{p}})(\mathbf{B}^{\prime}\cdot\hat{\mathbf{p}}) -i\frac{\eta  e v_F \hbar^2}{4c^2p^3} \mathbf{k}
(\mathbf{B}^{\prime}\cdot\mathbf{B}_{0,\lambda}) + i\frac{e\hbar^2}{4cp^3} \mathbf{k}(\mathbf{B}_{0,\lambda}\cdot[\mathbf{E}^{\prime}\times
\hat{\mathbf{p}}]),\\
&&\mathbf{v}^{\prime}=\frac{\lambda e v_F \hbar}{cp^2} \hat{\mathbf{p}} (\hat{\mathbf{p}}\cdot\mathbf{B}^{\prime})
-\frac{\lambda e v_F \hbar}{2cp^2} \mathbf{B}^{\prime} + \frac{5\eta  e^2 v_F \hbar^2}{8c^2 p^4} \hat{\mathbf{p}}(\hat{\mathbf{p}}\cdot
\mathbf{B}^{\prime})(\hat{\mathbf{p}}\cdot\mathbf{B}_{0,\lambda}) - \frac{\eta  e^2 v_F \hbar^2}{8c^2p^4}[\mathbf{B}^{\prime}(\hat{\mathbf{p}}
\cdot\mathbf{B}_{0,\lambda})+\mathbf{B}_{0,\lambda}(\mathbf{B}^{\prime}\cdot\hat{\mathbf{p}})] \nonumber\\
&&-\frac{3 \eta  e^2 \hbar^2 v_F }{4c^2p^4}\hat{\mathbf{p}}(\mathbf{B}^{\prime}\cdot\mathbf{B}_{0,\lambda})
+\frac{e^2\hbar^2}{4cp^4} (\mathbf{E}^{\prime}\times\mathbf{B}_{0,\lambda}) -\frac{e^2\hbar^2}{cp^4} \hat{\mathbf{p}}(\hat{\mathbf{p}}
\cdot[\mathbf{E}^{\prime}\times\mathbf{B}_{0,\lambda}]),\\
&&\kappa_{\lambda}\equiv \frac{e}{c} (\mathbf{\Omega}_{0,\lambda} \cdot \mathbf{B}_{0,\lambda})
= \lambda \eta  \hbar\frac{e\left(\hat{\mathbf{p}}\cdot\mathbf{B}_{0,\lambda}\right)}{2c p^2} +\frac{e^2 \hbar^2}{4c^2p^4}
\left[2(\hat{\mathbf{p}}\cdot\mathbf{B}_{0,\lambda})^2-B_{0,\lambda}^2\right].
\label{second-collective-formulas-ee}
\end{eqnarray}
By making use of the cylindrical coordinates (with the $z$ axis pointing along the magnetic field
$\mathbf{B}_{0}$ and $\phi$ being the azimuthal angle of momentum $\mathbf{p}$), we
rewrite Eq.~(\ref{second-collective-kinetic-equation})  in the following form:
\begin{eqnarray}
&&\zeta_{\lambda}
\frac{ev_F B_{0,\lambda}}{cp}\frac{\partial \delta f_{\lambda}}{\partial \phi} + i\left[(1+\kappa_{\lambda})\omega
- (\mathbf{k}\cdot\mathbf{v}_0)-\frac{e}{c}(\mathbf{v}_0\cdot\mathbf{\Omega}_{0,\lambda})(\mathbf{k}\cdot\mathbf{B}_{0,\lambda})\right]
\delta f_{\lambda} \nonumber\\
&&= \left[e(\tilde{\mathbf{E}}^{\prime}_{\lambda}\cdot\mathbf{v}_0) -\frac{ev_F}{c}\zeta_{\lambda}(\mathbf{v}^{\prime}
\cdot[\hat{\mathbf{p}}
\times\mathbf{B}_{0,\lambda}])
+\frac{e^2}{c}(\tilde{\mathbf{E}}_{\lambda}^{\prime}\cdot\mathbf{B}_{0,\lambda})(\mathbf{v}_0\cdot\mathbf{\Omega}_{0,\lambda}) \right]
\frac{\partial f_{\lambda}^{\rm (eq)}}{\partial \epsilon_\mathbf{p}},
\label{second-collective-kinetic-equation-1}
\end{eqnarray}
where we used the fact that
$\left(\mathbf{v}_0\times\mathbf{B}_{0,\lambda}\right)\propto\left(\mathbf{p}\times\mathbf{B}_{0,\lambda}\right)$, and introduced
the function
\begin{equation}
\zeta_{\lambda} \equiv \eta  \left\{1 +\frac{\lambda \eta  e \hbar (\hat{\mathbf{p}}\cdot\mathbf{B}_{0,\lambda})}{cp^2}
+\frac{5 e^2 \hbar^2 (\hat{\mathbf{p}}\cdot\mathbf{B}_{0,\lambda})^2}{16c^2 p^4} -\frac{3 e^2\hbar^2 B_{0,\lambda}^2}{8c^2p^4}\right\}.
\label{second-CKT-v-cross-B}
\end{equation}
In principle, the differential equation (\ref{second-collective-kinetic-equation-1}) for the oscillating part of the distribution function
$\delta f_{\lambda}(\phi)$ can be solved analytically. The corresponding analysis is rather tedious and will not be presented
here. In the next section, we will analyze a special case of longitudinal modes propagating along
the direction of the background magnetic and pseudomagnetic fields, i.e., $\mathbf{k}\parallel\mathbf{B}_{0,\lambda}$.

When the solution for $\delta f_{\lambda}$ is available, one can calculate all contributions
to the electric current [see Eqs.~(\ref{second-CKT-electric-current-b}), (\ref{second-consistent-4-current}),
and (\ref{second-consistent-current-density})], and then use Eq.~(\ref{second-polarization-tensor})
to determine the polarization vector. The formal result reads
\begin{eqnarray}
\mathbf{P}^{\prime} = \mathbf{P}^{\prime}_T+\mathbf{P}^{\prime (0)} +\mathbf{P}^{\prime (0)}_{M} +\mathbf{P}^{\prime (1)}
+\mathbf{P}^{\prime (1)}_{M},
\label{second-collective-kinetic-equation-P-def}
\end{eqnarray}
where
\begin{eqnarray}
\label{second-collective-kinetic-equation-P-be}
\mathbf{P}^{\prime}_T &=&i\frac{e^3}{2\pi^2 \hbar^2 c \omega}\,b_0 \,\mathbf{B}^{\prime}
- i\frac{e^3}{2\pi^2 \hbar^2 c \omega}\,(\mathbf{b}
\times\mathbf{E}^{\prime})
\end{eqnarray}
is the contribution due to the topological current in Eq.~(\ref{second-consistent-current-density}),
\begin{eqnarray}
\mathbf{P}^{\prime (0)}_{M}&=& \sum_{\eta=\pm} \eta\frac{ie}{\omega}\int\frac{d^3p}{(2\pi \hbar)^3} \left[\epsilon_{\mathbf{p}}(\partial_{\mathbf{r}}\times
\mathbf{\Omega}_{\lambda}^{\prime}) + (\partial_{\mathbf{r}}\times\epsilon_{\mathbf{p}}^{\prime}\mathbf{\Omega}_{0,\lambda}) \right]
f^{\rm (eq)}_{\lambda},\\
\mathbf{P}^{\prime (1)}_{M}&=&\sum_{\eta=\pm} \eta \frac{ie}{\omega}\int\frac{d^3p}{(2\pi \hbar)^3} \epsilon_{\mathbf{p}} (\partial_{\mathbf{r}}\times
\delta f_{\lambda}\mathbf{\Omega}_{0,\lambda}) =
-\sum_{\eta=\pm} \eta\frac{e}{\omega}\int\frac{d^3p}{(2\pi \hbar)^3} \epsilon_{\mathbf{p}}(\mathbf{k}\times\mathbf{\Omega}_{0,\lambda})\, \delta f_{\lambda}
\label{second-collective-kinetic-equation-P-ee}
\end{eqnarray}
are the contributions from the magnetization current, and
\begin{eqnarray}
\mathbf{P}^{\prime (0)}&=&\sum_{\eta=\pm} \eta \frac{i e}{\omega}\int\frac{d^3p}{(2\pi \hbar)^3}\left[\mathbf{v}^{\prime}
+\frac{e}{c}\mathbf{B}^{\prime}(\mathbf{\Omega}_{0,\lambda}\cdot \mathbf{v}_0)
+\frac{e}{c}\mathbf{B}_{0,\lambda}(\mathbf{\Omega}_{\lambda}^{\prime}\cdot \mathbf{v}_0)
+\frac{e}{c}\mathbf{B}_{0,\lambda}(\mathbf{\Omega}_{0,\lambda}\cdot \mathbf{v}^{\prime})
+e\left(\tilde{\mathbf{E}}_{\lambda}^{\prime}\times\mathbf{\Omega}_{0,\lambda}\right)\right]\,f_{\lambda}^{\rm (eq)},
\label{second-collective-kinetic-equation-P0prime}
\\
\mathbf{P}^{\prime (1)}&=&\sum_{\eta=\pm} \eta \frac{i e}{\omega}\int\frac{d^3p}{(2\pi \hbar)^3}\left[\mathbf{v}_0
+\frac{e}{c}\mathbf{B}_{0,\lambda}(\mathbf{\Omega}_{0,\lambda}\cdot \mathbf{v}_0)\right]\,\delta f_{\lambda}
\label{second-collective-kinetic-equation-P1prime}
\end{eqnarray}
are other contributions. Note that we used the following shorthand notations:
\begin{eqnarray}
\label{second-collective-formulas-1-be}
\mathbf{\Omega}_{\lambda}^{\prime} &=& \frac{e \hbar^2}{4p^4}\left\{\frac{2}{c}\hat{\mathbf{p}}(\hat{\mathbf{p}}\cdot\mathbf{B}^{\prime}) -
\frac{1}{c}\mathbf{B}^{\prime} +\frac{2\eta }{v_F}[\mathbf{E}^{\prime}\times\hat{\mathbf{p}}]\right\},\\
\epsilon_{\mathbf{p}}^{\prime} &=& -\frac{\lambda e v_F\hbar}{2cp} (\mathbf{B}^{\prime}\cdot\hat{\mathbf{p}}) +\frac{e^2\hbar^2}{4cp^3}
\left\{\frac{\eta  v_F}{2c}\left[2(\mathbf{B}^{\prime}\cdot\mathbf{B}_{0,\lambda}) -(\hat{\mathbf{p}}\cdot\mathbf{B}^{\prime})
(\hat{\mathbf{p}}\cdot\mathbf{B}_{0,\lambda})\right] -(\mathbf{B}_{0,\lambda}\cdot[\mathbf{E}^{\prime}\times\hat{\mathbf{p}}])\right\}.
\label{second-collective-formulas-1-ee}
\end{eqnarray}

\subsection{Chiral magnetic and chiral pseudomagnetic waves}
\label{sec:second-collective-CMW}

In a general case, the solution to Eq.~(\ref{second-collective-kinetic-equation-1}) is quite cumbersome. Here, for
simplicity, we will study only the collective modes propagating
along the direction of background magnetic and pseudomagnetic fields with $\mathbf{k}=k\hat{\mathbf{z}}$. It should be noted that, in this
case, the consistency of
Eqs.~(\ref{second-collective-B-tensor-spectrum-EM-1}) and (\ref{second-collective-kinetic-equation-P-be})
requires that the chiral shift $\mathbf{b}$ has no perpendicular component to the fields. In what follows,
therefore, we consider only the case $\mathbf{b}=b_{\parallel}\hat{\mathbf{z}}$. (For a discussion of the
effects of $\mathbf{b}_{\perp}$ on the collective excitations in the first-order theory, see
Refs.~\cite{Gorbar:2016ygi,Gorbar:2016sey}.)

Then, Eq.~(\ref{second-collective-kinetic-equation-1}) can be rendered in the following standard form
(see, e.g., Ref.~\cite{Landau:t10}):
\begin{equation}
\frac{\partial \delta f_{\lambda}}{\partial \phi} +i a\,\delta f_{\lambda}
=Q(\phi),
\label{second-collective-CMW-eq-general-2}
\end{equation}
where
\begin{eqnarray}
Q(\phi) &=& \frac{cp}{ev_F B_{0,\lambda}\zeta_{\lambda}}\left[e(\tilde{\mathbf{E}}^{\prime}_{\lambda}\cdot\mathbf{v}_0)
-\frac{ev_F}{c}\zeta_{\lambda}(\mathbf{v}^{\prime}\cdot[\hat{\mathbf{p}}\times\mathbf{B}_{0,\lambda}])
+\frac{e^2}{c}(\tilde{\mathbf{E}}_{\lambda}^{\prime}\cdot\mathbf{B}_{0,\lambda})(\mathbf{v}_0\cdot\mathbf{\Omega}_{0,\lambda}) \right]
\frac{\partial f_{\lambda}^{\rm (eq)}}{\partial \epsilon_\mathbf{p}},\\
a &=& \frac{cp}{ev_F B_{0,\lambda}\zeta_{\lambda}}\left[(1+\kappa_{\lambda})\omega - (\mathbf{k}\cdot\mathbf{v}_0)
-\frac{e}{c}kB_{0,\lambda}(\mathbf{v}_0\cdot\mathbf{\Omega}_{0,\lambda})\right].
\end{eqnarray}

The general solution to Eq.~(\ref{second-collective-CMW-eq-general-2}) reads as
\begin{equation}
\delta f_{\lambda} (\phi)  = C_1 e^{-ia \phi} + \int_{0}^{\phi-C_0} e^{-ia \tau} Q(\phi-\tau)  d\tau.
\label{second-collective-CMW-general-solution}
\end{equation}
For $\delta f_{\lambda}(\phi)$ to be a periodic function of $\phi$, one should set $C_1=0$ and $C_0\to \pm \infty $, where
the actual sign of $C_0$ is determined by $\eta\, \sign{eB_{0,\lambda}}$. Note that the latter choice ensures the finiteness
of the integral over $\tau$ in Eq.~(\ref{second-collective-CMW-general-solution}), provided $\omega\to\omega+i0$
inside the function $a$ in the exponent that mimics a gradual turning on of the perturbation fields \cite{Landau:t10}.
It is worth noting that the kinetic equation (\ref{second-collective-CMW-eq-general-2}) has no self-consistent
solutions for $\mathbf{E}^{\prime}=\mathbf{0}$. Indeed, if $\mathbf{E}^{\prime}=\mathbf{0}$, the Maxwell's
equations require that $\mathbf{B}^{\prime}=\mathbf{0}$ as well. Then, Eq.~(\ref{second-collective-CMW-eq-general-2})
reduces to a homogeneous equation, whose solution is given by the first term in Eq.~(\ref{second-collective-CMW-general-solution}).
However, this is not a valid solution since it cannot be periodic in $\phi$. We conclude, therefore, that the effects
of the dynamical electromagnetism are always present in the collective modes propagating along the external (pseudo)magnetic field,
including the CMW.

In the case of the CMW-type longitudinal modes, i.e., modes where the oscillating electric field is parallel to the external
(pseudo)magnetic field, i.e., $\mathbf{E}^{\prime}\parallel \hat{\mathbf{z}}$, function $Q$ does not depend on
the azimuthal angle $\phi$. An additional simplification arises from the absence of an oscillating magnetic field,
$\mathbf{B}^{\prime}=c[\mathbf{k}\times\mathbf{E}^{\prime}]/\omega=\mathbf{0}$, as follows from the Maxwell's equations.
Then the solution to Eq.~(\ref{second-collective-CMW-eq-general-2}) takes the following form:
\begin{equation}
\delta f_{\lambda} (\phi)=\delta f_{\lambda} = -i\frac{Q}{a}.
\label{second-collective-CMW-general-solution-1}
\end{equation}
With this solution at hand, we can now calculate the polarization vector by using the general expressions in
Eqs.~(\ref{second-collective-kinetic-equation-P-def})--(\ref{second-collective-kinetic-equation-P1prime}).
Performing the integrations over the polar and azimuthal angular coordinates, it is easy to see that
the only nontrivial contributions to the polarization vector (\ref{second-collective-kinetic-equation-P-def})
come from $P_{z}^{\prime (1)}$. After straightforward, although tedious,
calculations (see Appendix~\ref{sec:App-Polarization} for details), we obtain the following results:
\begin{equation}
\label{second-collective-CMW-P-ap-be}
P_{z}^{\prime (1)} \equiv \left(\chi_0^{33}+\chi_1^{33}+\chi_2^{33}\right)E_{z}^{\prime},
\end{equation}
where the three parts of the susceptibility tensor are given by
\begin{eqnarray}
\chi_0^{33} &=& \frac{3 n_0^2\Omega_e^2}{4\pi  v_F^2k^2} \left[1 -\frac{\omega}{2v_Fk}
\ln{\left(\frac{\omega+ v_Fk}{\omega- v_Fk}\right)}\right],\\
\chi_1^{33} &=& - \frac{e^3 v_F^2(\mathbf{B}_{0,5}\cdot\mathbf{k}) }{2\pi^2 c \hbar^2 \omega(\omega^2-v_F^2k^2)},\\
\chi_2^{33} &=& -\sum_{\lambda=\pm} \frac{1}{2T\Lambda_{\rm IR} \cosh^2{\left(\frac{\mu_{\lambda}}{2T}\right)}}
\frac{e^4 B_{0,\lambda}^2 }{192 c^2 \pi^2 \hbar v_F^3 k^5 (\omega^2-v_F^2k^2)^2}
\Bigg[2v_Fk\left(4v_F^6k^6-98\omega^2v_F^4k^4+229v_F^2k^2\omega^4-123\omega^6\right) \nonumber\\
&+&3\omega\left(41\omega^2-8v_F^2k^2\right)(\omega^2-v_F^2k^2)^2\ln{\left(\frac{\omega+ v_Fk}{\omega- v_Fk}\right)} \Bigg]
\nonumber\\
&+& \sum_{\lambda=\pm}\frac{1}{T}\frac{\partial F\left(\frac{\mu_{\lambda}}{T}\right)}{\partial \mu_{\lambda}}
\frac{e^4 B_{0,\lambda}^2}{192 c^2 \pi^2 \hbar v_F^2 k^5 (\omega^2-v_F^2k^2)^2}
\Bigg[6v_Fk \left(v_F^6k^6-35\omega^2v_F^4k^4+86\omega^4v_F^2k^2-48\omega^6\right) \nonumber\\
&+&6 \omega
\left(24\omega^2-3v_F^2k^2\right)(\omega^2-v_F^2k^2)^2\ln{\left(\frac{\omega+ v_Fk}{\omega- v_Fk}\right)} \Bigg].
\label{second-collective-CMW-P-ap-ee}
\end{eqnarray}
Here, $\Lambda_{\rm IR} = \sqrt{\hbar |eB_{0,\lambda}|/c}$ is an infrared cutoff and the function $F(x)$, as well as its Pad\'{e}
approximant are defined in Eqs.~(\ref{second-collective-CMW-F-def}) and (\ref{second-collective-CMW-F-Pade}), respectively. It is worth noting that the presence of this infrared singularity signifies that the expansion in $B_{0,\lambda}$ is nonperturbative. However, in the regime of small temperature $T\to0$ these terms are exponentially suppressed. Furthermore, we
introduced the shorthand notations for the coupling constant $\alpha=e^2/(\hbar v_Fn_0^2)$ and the Langmuir (plasma) frequency,
\begin{equation}
\Omega_e \equiv \sqrt{\frac{4\alpha}{3\pi\hbar^2}\left(\mu^2+\mu_5^2 +\frac{\pi^2 T^2}{3}\right)}.
\end{equation}
By making use of the polarization vector (\ref{second-collective-CMW-P-ap-be}), we rewrite the characteristic equation
(\ref{second-collective-B-tensor-dispersion-relation-general})
in the following form:
\begin{eqnarray}
\label{second-collective-CMW-disp-Eq}
n_0^2+4\pi{\left(\chi_0^{33}+\chi_1^{33}+\chi_2^{33} \right)} = 0.
\end{eqnarray}
In the limit of long wavelengths $ck\ll \Omega_e$ and small $B_{0}$ and $B_{0,5}$ fields, the
analytical solution to Eq.~(\ref{second-collective-CMW-disp-Eq}) reads
\begin{eqnarray}
\label{second-collective-CMW-app}
\omega &\simeq& \sqrt{\Omega_{e,B}^2 +\frac{2   \alpha e(\mathbf{B}_{0,5}\cdot\mathbf{k}) v_F^3}{\pi c\hbar \Omega_e} +A_1(v_Fk)^2 +
\frac{  \alpha e(\mathbf{B}_{0,5}\cdot\mathbf{k}) v_F^3}{5\pi c\hbar \Omega_e} \left(\frac{v_Fk}{\Omega_e}\right)^2 +A_2(v_Fk)^4\cdots},
\end{eqnarray}
where
\begin{eqnarray}
\label{plasma-gap}
\left(\Omega_{e,B}\right)^2 &\simeq & \Omega_e^2 +\frac{\alpha^2 v_F^4 \hbar n_{0}^2}{120\pi c^2 T^2} \sum_{\lambda=\pm}B_{0,\lambda}^2
\left[38\, T F_1(\mu_{\lambda},B_{0,\lambda}) -9v_F F_2(\mu_{\lambda})\right]
+O(B_{0}^3, B_{0,5}^3),\\
\label{A1}
A_{1} &\simeq& \frac{3}{5} - \frac{2\alpha^3 v_F^5n_0^2B_{0,5}^2}{\pi^2c^2\hbar \Omega_e^4} + \frac{\alpha^2 v_F^4 \hbar n_{0}^2}{700\pi c^2
\Omega_e^2 T^2} \sum_{\lambda=\pm}B_{0,\lambda}^2\Big[327\, T F_1(\mu_{\lambda},B_{0,\lambda}) -201v_F F_2(\mu_{\lambda}) \Big]+O(B_{0}^3,
B_{0,5}^3), \\
\label{A2}
A_{2} &\simeq& \frac{12}{175 \Omega_{e}^2} -\frac{4 \alpha^3 v_F^5 n_{0}^2B_{0,5}^2}{5\pi^2\hbar c^2 \Omega_e^6} +\frac{\alpha^2 v_F^4
\hbar n_0^2}{15750 \pi c^2 \Omega_e^4} \sum_{\lambda=\pm}B_{0,\lambda}^2\Big[5254\, T F_1(\mu_{\lambda},B_{0,\lambda}) -4767v_F F_2(\mu_{\lambda})
\Big] +O(B_{0}^3, B_{0,5}^3).\nonumber\\
\end{eqnarray}
Here, we also used the following functions:
\begin{eqnarray}
F_1\left(\frac{\mu_{\lambda}}{T},B_{0,\lambda}\right) &\equiv& \frac{1}{2\sqrt{\hbar|eB_{0,\lambda}|/c} \cosh^2{\left(\frac{\mu_{\lambda}}{2T} \right)}},
\label{eq55}
\\
F_2\left(\frac{\mu_{\lambda}}{T}\right) &\equiv&  T\frac{\partial F\left(\frac{\mu_{\lambda}}{T}\right)}{\partial \mu_{\lambda}}.
\label{eq56}
\end{eqnarray}
According to Eq.~(\ref{second-collective-CMW-app}), the frequency of the longitudinal mode depends linearly on the pseudomagnetic field
$\mathbf{B}_{0,5}$. It is natural to call the corresponding collective excitation in the presence of a strain-induced pseudomagnetic field
\emph{the chiral pseudomagnetic wave (CPMW)}. As we see from Eqs.~(\ref{second-collective-CMW-app}) and (\ref{plasma-gap}), both the CMW and CPMW are gapped plasmons.
Moreover, the values of their gaps contain corrections quadratic in the background magnetic and pseudomagnetic fields. There are also
quadratic corrections in the terms dependent on the wave vector [see Eqs.~(\ref{second-collective-CMW-app}), (\ref{A1}), and (\ref{A2})].

\begin{figure}[!t]
\begin{minipage}[ht]{0.45\linewidth}
\center{\includegraphics[width=1.0\linewidth]{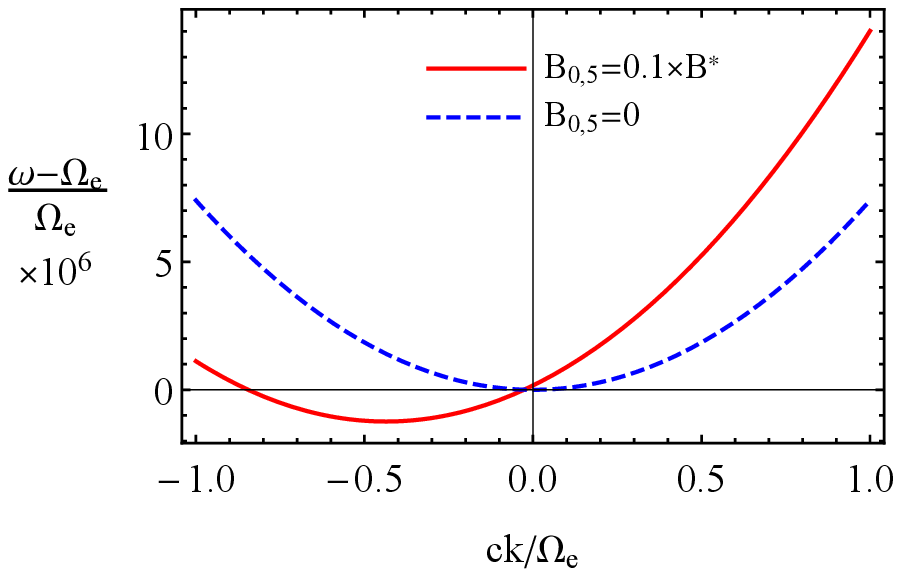} \\
{\small (a) $T=0$, $\mu_5=0$, $\mu=\sqrt{3\pi /(4 \alpha)}\hbar\Omega_e$}}
\end{minipage}
\begin{minipage}[ht]{0.45\linewidth}
\center{\includegraphics[width=1.0\linewidth]{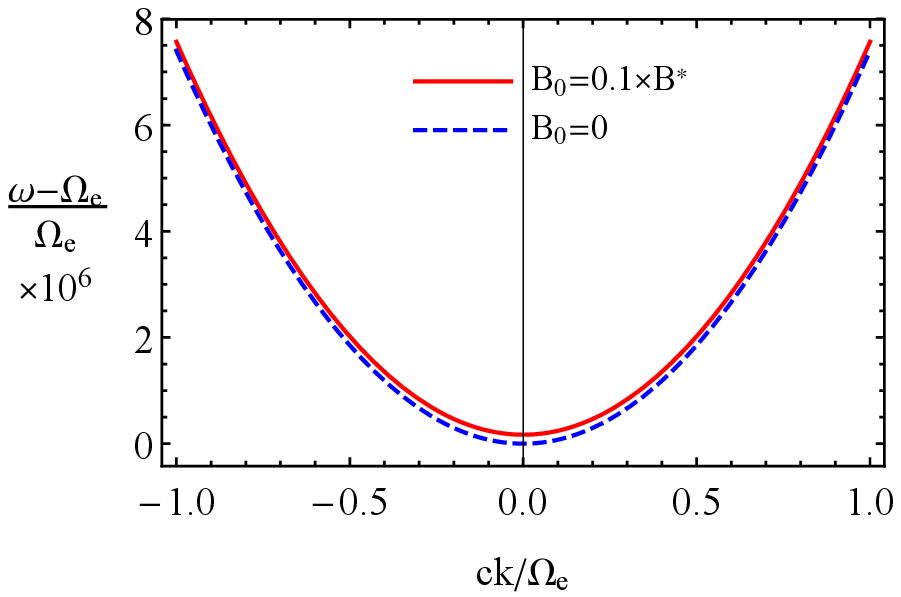} \\
{\small (b) $T=0$, $\mu_5=0$, $\mu=\sqrt{3\pi /(4 \alpha)}\hbar\Omega_e$}}
\end{minipage}
\begin{minipage}[ht]{0.45\linewidth}
\center{\includegraphics[width=1.0\linewidth]{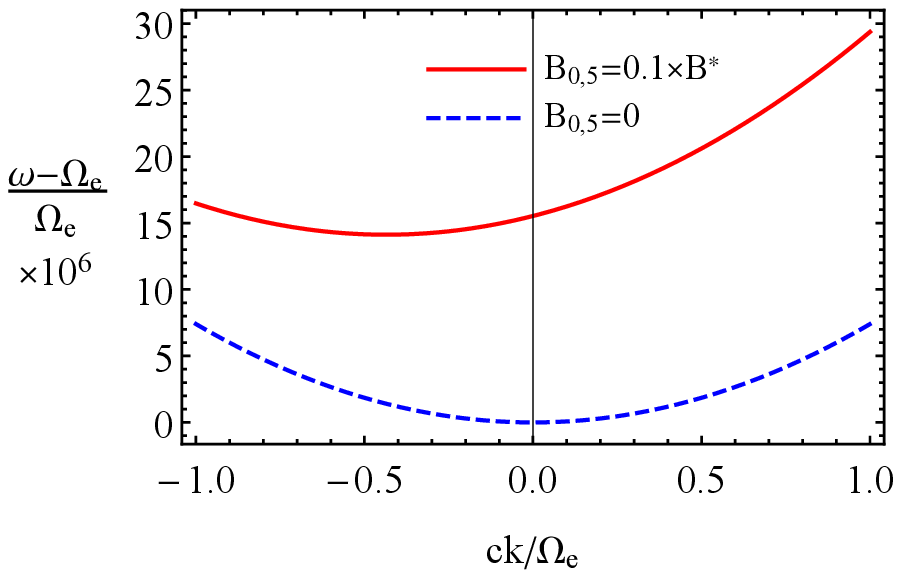} \\
{\small (c) $T=3\sqrt{1/(4\pi\alpha)}\hbar\Omega_e$, $\mu_5=0$, $\mu=0$}}
\end{minipage}
\begin{minipage}[ht]{0.45\linewidth}
\center{\includegraphics[width=1.0\linewidth]{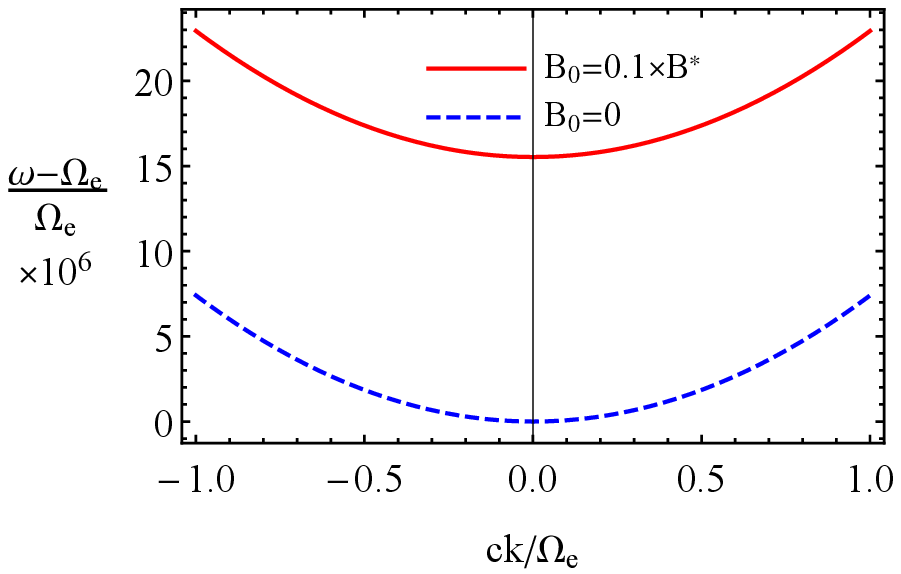} \\
{\small (d) $T=3\sqrt{1/(4\pi\alpha)}\hbar\Omega_e$, $\mu_5=0$, $\mu=0$}}
\end{minipage}
\caption{Left figures: The dispersion relation for the longitudinal mode propagating along the direction
of background magnetic and pseudomagnetic fields with the wave vector $\mathbf{k}=k\hat{\mathbf{z}}$
at $B_{0,5}=0.1B^{*}$ (red solid line) and $B_{0,5}=0$ (blue dashed line). Right figures: The same
dependence at $B_{0}=0.1B^{*}$ (red solid line) and $B_{0}=0$ (blue dashed line).}
\label{fig:second-collective-CMW-B-and-B5}
\end{figure}

The dispersion relations of the CMW and CPMW at different values of $T$, $\mu$, and $\mu_5$ are shown in
Fig.~\ref{fig:second-collective-CMW-B-and-B5}, where we use the following reference magnetic field:
\begin{equation}
B^{*} = \frac{c\hbar \Omega_e^2}{ev_F^2}.
\label{second-collective-CMW-B-ref}
\end{equation}

As we see from the left panel in Fig.~\ref{fig:second-collective-CMW-B-and-B5}, the dispersion relations for the CPMW have
minima at nonzero values of the wave vectors. Their approximate locations are determined by $ck/\Omega_e
\approx-5v_F\alpha eB_{0,5}/(3\pi \hbar \Omega_e^2)$.
We checked that interchanging the electric and chiral chemical potentials does not affect the properties of the CPMW.

Although the results in Fig.~\ref{fig:second-collective-CMW-B-and-B5} show that the dispersion relations of
the CMW and CPMW are indeed modified by the quadratic corrections in magnetic and pseudomagnetic fields, the
corresponding effect is weak at small temperatures. The correction to the plasma gap is much larger at
$T=3\sqrt{1/(4\pi\alpha)}\hbar\Omega_e$, $\mu_5=0$, and $\mu=0$, albeit it is still about five orders of
magnitude smaller that the Langmuir frequency.

As we emphasized in Refs.~\cite{Gorbar:2016ygi,Gorbar:2016sey}, the CMW and CPMW are chiral (pseudo)magnetic
plasmons. Their chiral nature is evident from the fact that both electric $J_z$  and chiral $J_z^5$ current densities
are oscillating in these
waves. This can be seen from the explicit expressions for $J_z$ and $J_z^5$, given by Eqs.~(\ref{second-collective-CMW-J})
and (\ref{second-collective-CMW-J-5}) in Appendix~\ref{sec:App-Polarization}, respectively. Leaving aside the technical
details, we would like to emphasize only that the chiral current density oscillates in space and time, i.e.,
$J_z^5\propto E_z\sin{(\omega t -\mathbf{k}\cdot\mathbf{r})}$, which is similar to the electric current density, i.e., $J_z\propto E_z\sin{(\omega t -\mathbf{k}\cdot\mathbf{r})}$. Note, however, that the amplitudes of the currents are different and depend on
the chiral chemical potential and the pseudomagnetic field.

\section{Summary}
\label{sec:Summary-Discussions}

By making use of the second-order corrections in (pseudo)electromagnetic fields to the quasiparticle
energy, as well as the first-order corrections to the Berry curvature, we derived a consistent chiral
kinetic theory valid to the second order in the electromagnetic and pseudoelectromagnetic fields
for the simplest model of relativistic matter with two Weyl fermions. Such an extended theory allows
one to study reliably various effects nonlinear in electromagnetic fields and is one of the main results of this paper.

While the semiclassical equations of motion preserve their form, the Berry curvature, as well as the quasiparticle
dispersion relation receive nontrivial field-dependent corrections. However, we found that,
even with the nontrivial corrections included,
the Berry curvature $\mathbf{\Omega}_{\lambda}$ still defines a
monopole-type vector field in the momentum space which corresponds to a unit of topological charge.
Therefore, the chiral anomaly relation retains its canonical form in the second-order formulation of
the chiral kinetic theory.

In order to illustrate the second-order consistent chiral kinetic theory, we analyzed the spectrum of longitudinal
collective modes propagating along the direction of background magnetic and pseudomagnetic fields.
We showed that, in the presence of a pseudomagnetic field, there is a different type of collective excitation
similar to the chiral magnetic wave, which we call \emph{the chiral pseudomagnetic wave}. The effects
of dynamical electromagnetism play an important role and transform the chiral magnetic and chiral
pseudomagnetic waves into plasmons with special properties. The latter manifest themselves in the
oscillations of the chiral current density, which are absent in the case of ordinary electromagnetic plasmons.

We found that the plasmon gaps receive corrections quadratic in magnetic and pseudomagnetic
fields, which cannot be reliably obtained within the conventional first-order chiral kinetic theory. Note, however,
that these corrections are estimated to be rather weak compared to the effects of dynamical
electromagnetism. In addition, the coefficients in front of odd powers of the wave vector in the dispersion
relation of the chiral pseudomagnetic wave are nonzero and proportional to the background pseudomagnetic
field. The coefficients in front of even powers depend on the square of both magnetic and pseudomagnetic fields.
Nevertheless, the leading contributions to these coefficients are due to the effects of dynamical electromagnetism.
It would be interesting to investigate
how these conclusions about the CMW and CPMW change in the case of strong background magnetic and/or
pseudomagnetic fields, and whether the effects of dynamical electromagnetism could be negligible. The corresponding problem
requires a framework that goes beyond the expansion in powers of $\mathbf{B}_0$ and $\mathbf{B}_{0,5}$ and, therefore,
will be considered elsewhere.

\begin{acknowledgments}
The work of E.V.G. was partially supported by the Program of Fundamental Research of the Physics and
Astronomy Division of the National Academy of Sciences of Ukraine.
The work of V.A.M. and P.O.S. was supported by the Natural Sciences and Engineering Research Council of Canada.
The work of I.A.S. was supported by the U.S. National Science Foundation under Grant No.~PHY-1404232.
\end{acknowledgments}

\appendix

\section{Corrections to the Berry curvature, the quasiparticle energy, and the velocity}
\label{sec:App-Corrections}

In this Appendix, we present the details of the derivation for the corrections to the Berry curvature, the
quasiparticle energy, and the velocity, which are needed in the formulation of the chiral kinetic theory
valid to the second order in (pseudo)electromagnetic fields.

Let us start from the derivation of the leading-order correction to the Berry curvature.
By making use of the formalism of Refs.~\cite{Gao:2014,Gao:2015}, the corresponding
correction has the following standard form:
\begin{equation}
\mathbf{\Omega}_{\lambda}^{(1)}= (\bm{\nabla}_{\mathbf{p}}\times\mathbf{a}^{\prime}_{\lambda}),
\label{second-def-Berry-curv-1App}
\end{equation}
where $\mathbf{a}^{\prime}_{\lambda}$ is a chirality ($\lambda=\pm$) and band-dependent ($\eta=\pm$) correction to the
Berry connection or positional shift. This correction is given by
\begin{equation}
\mathbf{a}^{\prime}_{\lambda}= \sum_{\xi \neq \eta } \left\{\frac{G_{\eta , \xi}\mathbf{A}_{\xi, \eta }}{\epsilon^{(0)}_{\eta }-
\epsilon^{(0)}_{\xi}} +
\frac{e}{4c} \partial_{\mathbf{p}_j} \left([\mathbf{B}_{\lambda}\times\mathbf{A}_{\eta , \xi}]_{j} \mathbf{A}_{\xi, \eta }\right)\right\} +
\frac{e\hbar^2}{v_F}\sum_{\xi\neq \eta } \frac{\mathbf{V}_{\eta  \xi}(\mathbf{V}_{\xi \eta }\cdot\mathbf{E}_{\lambda})}
{(\epsilon^{(0)}_{\eta }-\epsilon^{(0)}_{\xi})^3} + h.c.
\label{second-def-Berry-con-1App}
\end{equation}
Here, we used the interband matrix element associated with the magnetic dipole moment,
\begin{equation}
G_{\xi, \eta } =  -\frac{e}{2c}\left\{\sum_{\gamma\neq\eta} \left(\mathbf{B}_{\lambda}\cdot [\mathbf{V}_{\xi, \gamma}\times\mathbf{A}_{\gamma, \eta}]\right)
+ \left(\mathbf{B}_{\lambda}\cdot [\mathbf{V}_{\eta , \eta }\times\mathbf{A}_{\xi, \eta }]\right)\right\},
\end{equation}
which is defined in terms of the matrix elements of the velocity operator $\mathbf{V}=\lambda v_F\bm{\sigma}$ and the interband
Berry connection, i.e.,
\begin{eqnarray}
\mathbf{V}_{\xi, \eta }&=& \psi_{\xi}^{\dag} \mathbf{V} \psi_{\eta },\\
\mathbf{A}_{\xi, \eta }&=& -i\hbar\psi_{\xi}^{\dag} \bm{\partial}_{\mathbf{p}}\psi_{\eta },
\end{eqnarray}
respectively. Now, by making use of the normalized wave functions $\psi_{\eta }$ of the Weyl Hamiltonian (\ref{second-def-H})
\begin{eqnarray}
\psi_{\eta }&=& \sqrt{\frac{p +\eta \lambda p_z}{2p}}\left(
           \begin{array}{c}
             1 \\
             \frac{p_x+ip_y}{p_z+\eta \lambda p} \\
           \end{array}
         \right),
\label{second-def-psiApp}
\end{eqnarray}
we obtain the following explicit expressions for the correction to the Berry connection:
\begin{eqnarray}
\mathbf{a}^{\prime}_{\lambda}=\frac{e\hbar^2}{8p^3} \left\{ \frac{1}{c}[\mathbf{B}_{\lambda}\times\hat{\mathbf{p}}]
+\frac{2\eta }{v_F}\left[\mathbf{E}_{\lambda }-\hat{\mathbf{p}}(\hat{\mathbf{p}}\cdot\mathbf{E}_{\lambda})\right]\right\},
\label{second-def-Berry-con-3App}
\end{eqnarray}
where $\hat{\mathbf{p}}=\mathbf{p}/p$. Finally, by making use of Eq.~(\ref{second-def-Berry-curv-1App}), we
obtain the leading-order correction to the Berry curvature, i.e.,
\begin{equation}
\mathbf{\Omega}_{\lambda}^{(1)} = \frac{e \hbar^2}{4p^4} \left\{\frac{2}{c}\hat{\mathbf{p}}(\hat{\mathbf{p}}\cdot\mathbf{B}_{\lambda})
-\frac{1}{c}\mathbf{B}_{\lambda} +\frac{2\eta }{v_F}[\mathbf{E}_{\lambda}\times\hat{\mathbf{p}}]\right\}.
\label{def-Omega-1App}
\end{equation}

It remains to determine the quasiparticle dispersion relation to the second order in (pseudo)electromagnetic fields for the model at hand. The
corresponding general expression reads as follows \cite{Gao:2015}:
\begin{eqnarray}
\epsilon_{\mathbf{p}}= \epsilon^{(0)}_{\eta }- \frac{e}{c}(\mathbf{B}_{\lambda}\cdot\mathbf{m}) +\frac{e^2}{4c^2} (\mathbf{B}_{\lambda}\cdot
\mathbf{\Omega}_{\lambda}^{(0)}) (\mathbf{B}_{\lambda}\cdot\mathbf{m}) -\frac{e}{c}(\mathbf{B}_{\lambda} \cdot [\mathbf{a}_{\lambda}^{\prime}
\times\mathbf{V}_{\eta , \eta }]) +(\bm{\nabla}_{\mathbf{p}}\cdot \mathbf{P}_{E}) +\sum_{\xi \neq \eta } \frac{G_{\eta  \xi}G_{\xi, \eta }}
{\epsilon^{(0)}_{\eta }-\epsilon^{(0)}_{\xi}},
\label{second-energy-def}
\end{eqnarray}
where the orbital magnetic moment equals
\begin{equation}
\mathbf{m}= \frac{1}{2}\mathrm{Im} \left[ (\bm{\partial}_{\mathbf{p}}\psi_{\eta })^{\dag}\times \left(\epsilon^{(0)}_{\eta }
-\hat{H}\right)(\bm{\partial}_{\mathbf{p}}\psi_{\eta }) \right] =\epsilon^{(0)}_{\eta } \mathbf{\Omega}_{\lambda}^{(0)}
\end{equation}
and
\begin{equation}
\mathbf{P}_E = \frac{e^2\hbar^2}{4c^2} \left\{ ([\mathbf{B}_{\lambda}\times \bm{D}] \psi_{\eta })^{\dag} (\mathbf{V}
+\mathbf{V}_{\eta , \eta })
([\mathbf{B}_{\lambda}\times\bm{D}] \psi_{\eta }) +h.c.\right\}
\end{equation}
is the energy polarization density. Here, $\mathbf{D}= \hbar\bm{\partial}_{\mathbf{p}} -i\mathbf{A}_{\eta , \eta }$
is the gauge covariant derivative. After lengthy but straightforward calculations, we found that $\mathbf{P}_E =\mathbf{0}$ for
the Weyl Hamiltonian (\ref{second-def-H}). Moreover, one can check that all off-diagonal interband mixing terms vanish,
i.e., $G_{-\eta , \eta }=0$. The fourth term in Eq.~(\ref{second-energy-def}) reads as
\begin{eqnarray}
\label{second-energy-cor-be}
-\frac{e}{c}\left(\mathbf{B}_{\lambda} \cdot [\mathbf{a}_{\lambda}^{\prime}\times\mathbf{V}_{\eta , \eta }]\right) &=&
\frac{e^2 \hbar^2}{4cp^4} \left\{\frac{\eta  v_F}{2p c} \left[B_{\lambda}^2p^2-(\mathbf{B}_{\lambda}\cdot\mathbf{p})^2\right]
- \left(\mathbf{B}_{\lambda}\cdot[\mathbf{E}_{\lambda}\times\mathbf{p}]\right)\right\}.
\label{second-energy-cor-ee}
\end{eqnarray}
Thus, one finds that the quasiparticle dispersion relation (\ref{second-energy-def}) to the second order in (pseudo)electromagnetic fields equals
\begin{eqnarray}
\epsilon_{\mathbf{p}}\equiv \epsilon^{(0)}_{\eta }+\epsilon_{\eta }^{(1)}+\epsilon_{\eta }^{(2)} = \eta  v_Fp
-\lambda\frac{e\hbar v_F }{2cp} (\mathbf{B}_{\lambda}\cdot\hat{\mathbf{p}}) + \frac{e^2 \hbar^2}{4cp^3} \left\{\frac{\eta  v_F}{4c}
\left[2 B_{\lambda}^2-(\mathbf{B}_{\lambda}\cdot\hat{\mathbf{p}})^2\right] - (\mathbf{B}_{\lambda}\cdot[\mathbf{E}_{\lambda}\times
\hat{\mathbf{p}}])\right\},
\label{second-energy-2App}
\end{eqnarray}
where
\begin{eqnarray}
\label{second-energy-2-coef-be}
\epsilon_{\eta }^{(0)}&=&\eta  v_Fp, \\
\label{second-energy-2-coef-1}
\epsilon_{\eta }^{(1)}&=& - \eta \frac{ev_Fp}{c}(\mathbf{B}_{\lambda}\cdot\mathbf{\Omega}_{\lambda}^{(0)})
= -\lambda\frac{e\hbar v_F }{2cp} (\mathbf{B}_{\lambda}\cdot\hat{\mathbf{p}}), \\
\epsilon_{\eta }^{(2)}&=&\eta \frac{e^2v_Fp}{4c^2}(\mathbf{B}_{\lambda}\cdot\mathbf{\Omega}_{\lambda}^{(0)})^2
+ \frac{e^2 \hbar^2}{4cp^4} \left\{\frac{\eta  v_F}{2p c} \left[B_{\lambda}^2p^2-(\mathbf{B}_{\lambda}\cdot\mathbf{p})^2\right]
- \left(\mathbf{B}_{\lambda}\cdot[\mathbf{E}_{\lambda}\times\mathbf{p}]\right)\right\} \nonumber\\
&=& \frac{e^2 \hbar^2}{4cp^3} \left\{\frac{\eta  v_F}{4c} \left[2 B_{\lambda}^2-(\mathbf{B}_{\lambda}\cdot\hat{\mathbf{p}})^2\right]
- (\mathbf{B}_{\lambda}\cdot[\mathbf{E}_{\lambda}\times\hat{\mathbf{p}}])\right\}.
\label{second-energy-2-coef-ee}
\end{eqnarray}

The corresponding quasiparticle velocity is given by
\begin{equation}
\mathbf{v}=\frac{\partial \epsilon_{\mathbf{p}}}{\partial \mathbf{p}}=\mathbf{v}^{(0)}+\mathbf{v}^{(1)}+\mathbf{v}^{(2)},
\label{second-energy-v}
\end{equation}
where
\begin{eqnarray}
\label{second-energy-v-coef-be}
\mathbf{v}^{(0)}&=&\eta  v_F\hat{\mathbf{p}}, \\
\mathbf{v}^{(1)}&=&\lambda \frac{ev_F\hbar}{c} \hat{\mathbf{p}}\frac{(\hat{\mathbf{p}} \cdot\mathbf{B}_{\lambda})}{p^2}
-\lambda \frac{ev_F\hbar}{c} \frac{\mathbf{B}_{\lambda}}{2p^2},\\
\mathbf{v}^{(2)}&=& \frac{5\eta  e^2v_F \hbar^2 \hat{\mathbf{p}} (\mathbf{B}_{\lambda}\cdot\hat{\mathbf{p}})^2}{16c^2p^4}
-\frac{\eta e^2v_F \hbar^2 \mathbf{B}_{\lambda}(\mathbf{B}_{\lambda}\cdot\hat{\mathbf{p}})}{8c^2p^4}
- \frac{3\eta  e^2 \hbar^2v_F}{8c^2p^4}\hat{\mathbf{p}}B_{\lambda}^2 +\frac{e^2 \hbar^2}{4cp^4 }(\mathbf{E}_{\lambda}\times
\mathbf{B}_{\lambda})
+\frac{e^2 \hbar^2}{cp^4}\hat{\mathbf{p}}(\mathbf{B}_{\lambda}\cdot[\mathbf{E}_{\lambda}\times\hat{\mathbf{p}}]). \nonumber\\
\label{second-energy-v-coef-ee}
\end{eqnarray}

\section{Polarization vector and currents}
\label{sec:App-Polarization}

In this Appendix, we provide the details of the calculation of the polarization vector (\ref{second-collective-kinetic-equation-P-def}) and present an explicit form of the electric and chiral currents for the chiral magnetic and pseudomagnetic waves. Using
Eqs.~(\ref{second-collective-kinetic-equation-P-be}) through (\ref{second-collective-kinetic-equation-P-ee}) from the main text and integrating
over polar and azimuthal angles, one can show that $\mathbf{P}^{\prime (0)}=\mathbf{P}_{M}^{\prime (0)}=\mathbf{P}_{M}^{\prime (1)}=\mathbf{0}$ and
$P_{x}^{\prime (1)}=P_{y}^{\prime (1)}=0$. Moreover, for $\mathbf{b}\parallel\hat{\mathbf{z}}$, the topological part of the polarization vector (\ref{second-collective-kinetic-equation-P-be}) is also absent, i.e., $\mathbf{P}_{T}=\mathbf{0}$. However, the component of the polarization vector along the direction of (pseudo)magnetic field
$\hat{\mathbf{z}}$ is nonzero and equals
\begin{equation}
P_{z}^{\prime (1)} = \left(\chi_0^{33}+\chi_1^{33}+\chi_2^{33}\right)E_{z}^{\prime}.
\end{equation}

Components of the susceptibility tensor $\chi^{33}$ equal
\begin{eqnarray}
\label{second-collective-CMW-P-be}
\chi_0^{33} &=&-\sum_{\lambda=\pm} \sum_{\eta=\pm} \eta\int dp\, p^2 \frac{e^2}{2\pi^2 \hbar^3 k^2} \left[1 -\eta \frac{\omega}{2v_Fk}
\ln{\left(\frac{\eta \omega+v_Fk}{\eta \omega- v_Fk}\right)}\right] f_1 , \\
\chi_1^{33} &=& -\sum_{\lambda=\pm}\sum_{\eta=\pm} \eta \int dp \frac{e^3 \eta \lambda B_{0,\lambda}}{24\pi^2 c \hbar^2 v_F^2 k^4 (\omega^2-v_F^2k^2)} \Big\{2\eta  v_Fk
\left[-v_F^4k^4(2f_1-f_2)+2\omega^2v_F^2k^2(f_1+f_2)-3\omega^4(f_1+f_2)\right] \nonumber\\
&+&3\omega^3(\omega^2-v_F^2k^2)(f_1+f_2)\ln{\left(\frac{\eta \omega+ v_Fk}{\eta \omega- v_Fk}\right)}\Big\},\\
\chi_2^{33} &=& \sum_{\lambda=\pm}\sum_{\eta=\pm} \eta \int \frac{dp}{p^2}  \frac{\eta  e^4 B_{0,\lambda}^2}{192\pi^2 c^2 \hbar v_F^3k^5 (\omega^2-v_F^2k^2)^2}
\Bigg\{2\eta  v_Fk \Big[v_F^6k^6(4f_1-f_2-2f_3) -\omega^2v_F^4k^4(98f_1+7f_2+2f_3) \nonumber\\
&+&\omega^4v_F^2k^2(229f_1+29f_2+10f_3)-3\omega^6(41f_1+7f_2+2f_3)\Big] \nonumber\\
&-&3\omega(\omega^2-v_F^2k^2)^2\left[2v_F^2k^2(4f_1-f_2)
-\omega^2(41f_1+7f_2+2f_3)\right]\ln{\left(\frac{\eta \omega+ v_Fk}{\eta \omega- v_Fk}\right)}\Bigg\},
\label{second-collective-CMW-P-ee}
\end{eqnarray}
where
\begin{eqnarray}
f_1&=&\frac{\partial f^{(0)}_{\lambda} }{\partial \epsilon_{\mathbf{p}}},\\
f_2&=&\eta  v_Fp\frac{\partial^2 f^{(0)}_{\lambda}}{\partial \epsilon_{\mathbf{p}}^2},\\
f_3&=&(v_Fp)^2\frac{\partial^3 f^{(0)}_{\lambda}}{\partial \epsilon_{\mathbf{p}}^3},
\end{eqnarray}
and we used the following integrals:
\begin{eqnarray}
\label{second-collective-CMW-ints-be}
\int dp\, p^2 \frac{\partial f^{(0)}_{\lambda}}{\partial \epsilon_{\mathbf{p}}} &=& \eta\frac{2T^2}{v_F^3} \mbox{Li}_{2}
\left(-e^{\eta \mu_{\lambda}/T}\right),\\
\int dp\, \frac{\partial f^{(0)}_{\lambda}}{\partial \epsilon_{\mathbf{p}}} &=& -\frac{\eta}{v_F} \frac{1}{1+e^{-\eta \mu_{\lambda}/T}},\\
\int dp\, p \frac{\partial^2 f^{(0)}_{\lambda}}{\partial \epsilon_{\mathbf{p}}^2} &=&\frac{1}{v_F^2}
\frac{1}{1+e^{-\eta \mu_{\lambda}/T}},\\
\int \frac{dp}{p^2} \frac{\partial f^{(0)}_{\lambda}}{\partial \epsilon_{\mathbf{p}}} &=& \eta v_F\Bigg\{ -\frac{1}{\eta  v_Fp}
\frac{\partial f^{(0)}_{\lambda}}{\partial \epsilon_{\mathbf{p}}} \Big|^{\infty}_{\Lambda_{\rm IR}}
+\int_{\Lambda_{\rm IR}}^{\infty}\frac{dp}{p}
\frac{\partial^2 f^{(0)}_{\lambda}}{\partial \epsilon_{\mathbf{p}}^2} \Bigg\} = -\frac{\eta}{4T\Lambda_{\rm IR}
\cosh^2{\left(\frac{\mu_{\lambda}}{2T}\right)}}
+\eta  v_F\int_{\Lambda_{\rm IR}}^{\infty}\frac{dp}{p} \frac{\partial^2 f^{(0)}_{\lambda}}{\partial \epsilon_{\mathbf{p}}^2},\\
\int \frac{dp}{p} \frac{\partial^2 f^{(0)}_{\lambda}}{\partial \epsilon_{\mathbf{p}}^2} &=& \int_{\Lambda_{\rm IR}}^{\infty}\frac{dp}{p}
\frac{\partial^2 f^{(0)}_{\lambda}}{\partial \epsilon_{\mathbf{p}}^2},\\
\int dp \frac{\partial^3 f^{(0)}_{\lambda}}{\partial \epsilon_{\mathbf{p}}^3} &=& \frac{\eta}{4v_FT^2}
\frac{\tanh{\left(\frac{\eta  \mu_{\lambda}}{2T}\right)}}{\cosh^2{\left(\frac{\mu_{\lambda}}{2T}\right)}}.
\label{second-collective-CMW-ints-ee}
\end{eqnarray}
Here, the equilibrium distribution function was expanded as
\begin{equation}
f_{\lambda}^{\rm (eq)}\simeq f_{\lambda}^{(0)} +\left(\epsilon^{(1)}_{\eta }+\epsilon^{(2)}_{\eta}\right)
\frac{\partial f^{(0)}_{\lambda}}{\partial \epsilon_{\mathbf{p}}}  +\frac{\left(\epsilon^{(1)}_{\eta }\right)^2}{2}
\frac{\partial^2 f^{(0)}_{\lambda}}{\partial \epsilon_{\mathbf{p}}^2} + O(B_{0,\lambda}^3),
\end{equation}
where $f_{\lambda}^{(0)}$ is given by Eq.~(\ref{second-CKT-equilibrium-function}) with $\epsilon_{\mathbf{p}}=\epsilon^{(0)}_{\eta}$. While the quasiparticle energy without electromagnetic fields $\epsilon^{(0)}_{\eta}$ is given by Eq.~(\ref{second-energy-2-coef-be}), the field-dependent corrections $\epsilon^{(1)}_{\eta}$ and $\epsilon^{(2)}_{\eta}$ are given by Eqs.~(\ref{second-energy-2-coef-1}) and
(\ref{second-energy-2-coef-ee}), respectively. Next, we introduced an infrared cutoff $\Lambda_{\rm IR}=C\sqrt{\hbar |eB_{0,\lambda}|/c}$ with a
numerical constant $C$ of order unity. Such a cutoff has a transparent physical meaning: It separates the phase space
of large momenta, where the semiclassical description is valid, from the infrared region $p\lesssim \Lambda_{\rm IR}$,
where such a description fails (for details, see also Ref.~\cite{Stephanov:2012ki}). In our numerical calculations, we will use
$C=1$. After adding the contribution of antiparticles, one can set $\Lambda_{\rm IR}$ to zero in the last term of the
fourth integral and in the fifth integral because the corresponding integrals are no longer divergent in
the infrared region and can be expressed in terms of the derivative of the function
\begin{eqnarray}
\label{second-collective-CMW-F-def}
F\left(\nu_\lambda \right) \equiv -T\sum_{\eta=\pm} \int\frac{dp}{p}  \frac{\partial f^{(0)}_{\lambda}}{\partial \epsilon_{\mathbf{p}}}
\end{eqnarray}
with respect to $\nu_\lambda \equiv \mu_{\lambda}/T$, i.e.,
\begin{equation}
\frac{1}{T^2} \frac{\partial F\left(\nu_{\lambda}\right)}{\partial \nu_{\lambda}}=\sum_{\eta=\pm} \eta\int_{0}^{\infty}\frac{dp}{p} \frac{\partial^2 f^{(0)}_{\lambda}}{\partial \epsilon_{\mathbf{p}}^2}.
\end{equation}
High- and low-temperature asymptotes of $F(\nu_\lambda)$ equal
$F\left(\nu_\lambda\right)\simeq 7 \zeta(3) \nu_\lambda/(2\pi^2)\approx 0.426  \nu_\lambda$ for $T\to \infty$ and
$F\left(\nu_\lambda\right)\simeq \nu_\lambda^{-1}$ for $T\to 0$, respectively. The function $F\left(\nu_\lambda \right)$ could be
well approximated by the Pad\'e approximant of order [5/6], i.e.,
\begin{eqnarray}
\label{second-collective-CMW-F-Pade}
F\left(\nu_\lambda \right) \simeq \frac{7\zeta(3)}{2\pi^2}  \frac{\nu_\lambda +0.03533\nu_\lambda^3 +0.0007432\nu_\lambda^5}
{1+0.2290\nu_\lambda^2+0.01567 \nu_\lambda^4+0.0003098\nu_\lambda^6}.
\end{eqnarray}

Using Eqs.~(\ref{second-collective-CMW-ints-be})-(\ref{second-collective-CMW-ints-ee}), we obtain
\begin{eqnarray}
\label{second-collective-CMW-P-ap-1-be}
{\chi_0^{33}} &=& -\sum_{\eta=\pm}\sum_{\lambda=\pm}\frac{e^2}{2\pi^2 \hbar^3 k^2} \left[1 -\eta \frac{\omega}{2v_Fk}
\ln{\left(\frac{\eta \omega+ v_Fk}{\eta \omega- v_Fk}\right)}\right] \frac{T^2}{v_F^3}
\mbox{Li}_{2}\left(-e^{\eta  \mu_{\lambda}/T}\right),\\
{\chi_1^{33}} &=& - \sum_{\eta=\pm}\sum_{\lambda=\pm}\frac{e^3 \lambda B_{0,\lambda} v_F^2k}{4\pi^2 c
\hbar^2 \omega(\omega^2-v_F^2k^2)} \frac{1}{1+e^{-\eta  \mu_{\lambda}}},\\
{\chi_2^{33}} &=& -\sum_{\eta=\pm}\sum_{\lambda=\pm} \frac{1}{4T\Lambda_{\rm IR} \cosh^2{\left(\frac{\mu_{\lambda}}{2T}\right)}}
\frac{e^4B_{0,\lambda}^2 }{192 c^2 \pi^2 \hbar v_F^3 k^5 (\omega^2-v_F^2k^2)^2} \Bigg[2v_Fk
\left(4v_F^6k^6-98\omega^2v_F^4k^4+229\omega^4v_F^2k^2-123\omega^6\right) \nonumber\\
&+&3\eta \omega\left(41\omega^2-8v_F^2k^2\right)(\omega^2-v_F^2k^2)^2\ln{\left(\frac{\eta \omega+ v_Fk}{\eta \omega- v_Fk}\right)} \Bigg]
\nonumber \\
&+&\sum_{\eta=\pm} \sum_{\lambda=\pm}\frac{1 }{T}\frac{\partial F\left(\frac{\mu_{\lambda}}{T}\right)}{\partial \mu_{\lambda}}
\frac{e^4 B_{0,\lambda}^2}{192 c^2 \pi^2 \hbar v_F^2 k^5 (\omega^2-v_F^2k^2)^2} \Bigg[6v_Fk
\left(v_F^6k^6-35\omega^2v_F^4k^4+86\omega^4v_F^2k^2-48\omega^6\right) \nonumber\\
&+&6\eta \omega\left(24\omega^2-3v_F^2k^2\right)
(\omega^2-v_F^2k^2)\ln{\left(\frac{\eta \omega+ v_Fk}{\eta \omega- v_Fk}\right)} \Bigg]\nonumber\\
&-&\sum_{\eta=\pm}\sum_{\lambda=\pm} \frac{1}{4v_FT^2} \frac{\tanh{\left(\frac{\eta \mu_{\lambda}}{2T}\right)}}
{\cosh^2{\left(\frac{\mu_{\lambda}}{2T}\right)}} \frac{e^4 B_{0,\lambda}^2}{192 c^2 \pi^2 \hbar v_F k^5}
\Bigg[4v_F k(3\omega^2+v_F^2k^2)-6\eta \omega^3\ln{\left(\frac{\eta \omega+ v_Fk}{\eta \omega- v_Fk}\right)}\Bigg].
\label{second-collective-CMW-P-ap-1-ee}
\end{eqnarray}

Further, let us present explicit expressions for the nonzero components of the electric $J_z$ and chiral $J_z^5$ current densities,
i.e.,
\begin{eqnarray}
J_z&=&E_{z} \sin{(\omega t -\mathbf{k}\cdot\mathbf{r})} \frac{3\omega n_0^2 \Omega_e^2}{4\pi v_F^2k^2} \left[1 - \frac{\omega}{2v_Fk}
\ln{\left(\frac{ \omega+ v_Fk}{ \omega- v_Fk}\right)}\right]\nonumber\\
&+&E_{z} \sin{(\omega t -\mathbf{k}\cdot\mathbf{r})} \frac{\alpha n_0^2 e(\mathbf{B}_{0,5}\cdot\mathbf{k}) v_F^3}{2\pi^2 c
\hbar (\omega^2-v_F^2k^2)}\nonumber\\
&+&E_{z} \sin{(\omega t -\mathbf{k}\cdot\mathbf{r})} \sum_{\lambda=\pm} \frac{1}{T} F_1\left(\frac{\mu_{\lambda}}{T},B_{0,\lambda}\right)
\frac{\alpha^2 \hbar n_0^4 B_{0,\lambda}^2 \omega}{192 c^2 \pi^2 v_F k^5 (\omega^2-v_F^2k^2)^2} \Bigg[2v_Fk
\left(4v_F^6k^6-98\omega^2v_F^4k^4+229v_F^2k^2\omega^4-123\omega^6\right) \nonumber\\
&+&3 \omega\left(41\omega^2-8v_F^2k^2\right)(\omega^2-v_F^2k^2)^2\ln{\left(\frac{ \omega+ v_Fk}{ \omega- v_Fk}\right)} \Bigg]
\nonumber \\
&-& E_{z} \sin{(\omega t -\mathbf{k}\cdot\mathbf{r})}\sum_{\lambda=\pm}\frac{1}{T^2}F_{2}\left(\frac{\mu_{\lambda}}{T}\right) \frac{\alpha^2
\hbar n_0^4 B_{0,\lambda}^2 \omega}{192 c^2 \pi^2 k^5 (\omega^2-v_F^2k^2)^2} \Bigg[6v_Fk
\left(v_F^6k^6-35\omega^2v_F^4k^4+86\omega^4v_F^2k^2-48\omega^6\right) \nonumber\\
&+&6 \omega\left(24\omega^2-3v_F^2k^2\right)
(\omega^2-v_F^2k^2)^2\ln{\left(\frac{ \omega+ v_Fk}{ \omega- v_Fk}\right)} \Bigg]
\label{second-collective-CMW-J}
\end{eqnarray}
and
\begin{eqnarray}
J_z^{5}&=&E_{z} \sin{(\omega t -\mathbf{k}\cdot\mathbf{r})} \frac{2\alpha n_0^2 \omega \mu\mu_{5}}{\pi^2 \hbar^2 v_F^2k^2} \left[1 -
\frac{\omega}{2v_Fk}\ln{\left(\frac{ \omega+ v_Fk}{ \omega- v_Fk}\right)}\right] \nonumber\\
&+&E_{z} \sin{(\omega t -\mathbf{k}\cdot\mathbf{r})} \frac{\alpha n_0^2 e(\mathbf{B}_{0}\cdot\mathbf{k}) v_F^3}{2\pi^2 c \hbar
(\omega^2-v_F^2k^2)}\nonumber\\
&+&E_{z} \sin{(\omega t -\mathbf{k}\cdot\mathbf{r})} \sum_{\lambda=\pm} \frac{\lambda}{T} F_1\left(\frac{\mu_{\lambda}}{T},B_{0,\lambda}\right)
\frac{\alpha^2 \hbar n_0^4 B_{0,\lambda}^2 \omega}{192 c^2 \pi^2 v_F k^5 (\omega^2-v_F^2k^2)^2} \Bigg[2v_Fk
\left(4v_F^6k^6-98\omega^2v_F^4k^4+229v_F^2k^2\omega^4-123\omega^6\right) \nonumber\\
&+&3 \omega\left(41\omega^2-8v_F^2k^2\right)(\omega^2-v_F^2k^2)^2\ln{\left(\frac{ \omega+ v_Fk}{ \omega- v_Fk}\right)} \Bigg]
\nonumber \\
&-& E_{z} \sin{(\omega t -\mathbf{k}\cdot\mathbf{r})}\sum_{\lambda=\pm} \frac{\lambda}{T^2}F_{2}\left(\frac{\mu_{\lambda}}{T}\right)
\frac{\alpha^2 \hbar n_0^4 B_{0,\lambda}^2 \omega}{192 c^2 \pi^2 k^5 (\omega^2-v_F^2k^2)^2} \Bigg[6v_Fk
\left(v_F^6k^6-35\omega^2v_F^4k^4+86\omega^4v_F^2k^2-48\omega^6\right) \nonumber\\
&+&6 \omega\left(24\omega^2-3v_F^2k^2\right)
(\omega^2-v_F^2k^2)^2\ln{\left(\frac{ \omega+ v_Fk}{ \omega- v_Fk}\right)} \Bigg],
\label{second-collective-CMW-J-5}
\end{eqnarray}
respectively. Here, functions $F_1(x,y)$ and $F_2(x)$ are given by Eqs.~(\ref{eq55}) and (\ref{eq56}) in the main text. As one can see from the above expressions, both electric and chiral current densities are oscillating in the chiral magnetic and pseudomagnetic waves, albeit with different amplitudes.

\end{document}